\documentclass{article}
\usepackage[a4paper, total={6in, 8in}]{geometry}
\usepackage[utf8]{inputenc}
\usepackage{graphicx}
\usepackage{xcolor}
\usepackage{subcaption}
\usepackage{wrapfig}

\usepackage{setspace}
\usepackage{lineno}

\usepackage{natbib}
\title{Significant changes in EEG neural oscillations during different phases of three-dimensional multiple object tracking task (3D-MOT) imply different roles for attention and working memory.}

\author{Yannick Roy, Jocelyn Faubert}

\begin{document}

\maketitle

\textbf{Authors Information}\\
\\
Address: Faubert Lab, Université de Montréal, H3T 1P1, Québec, Canada \\
Corresponding author: yannick.roy@umontreal.ca  \\
Data and Code available: https://github.com/royyannick/3DMOT\_EEG  \\





\begin{abstract}
Our ability to track multiple objects in a dynamic environment enables us to perform everyday tasks such as driving, playing team sports, and walking in a crowded mall. Despite more than three decades of literature on multiple object tracking (MOT) tasks, the underlying and intertwined neural mechanisms remain poorly understood. Here we looked at the electroencephalography (EEG) neural correlates and their changes across the three phases of a 3D-MOT task, namely identification, tracking and recall. We recorded the EEG activity of 24 participants while they were performing a 3D-MOT task with either 1, 2 or 3 targets where some trials were lateralized and some were not. We observed what seems to be a handoff between focused attention and working memory processes when going from tracking to recall. Our findings revealed a strong inhibition in delta and theta frequencies from the frontal region during tracking, followed by a strong (re)activation of these same frequencies during recall. Our results also showed contralateral delay activity (CDA) for the lateralized trials, in both the identification and recall phases but not during tracking.
\end{abstract}

\section{Significance Statement}
Multiple Object Tracking (MOT) is heavily studied both in human cognition and computer vision. While MOT tasks have been used extensively in research and even commercially as a cognitive training tool, our understanding of the roles and relationship between attention and working memory remains vague. There is no clear consensus yet as how separated or intertwined these two mechanisms really are. Here, by looking at the brain activity both in the time domain and frequency domain, we show a clear cognitive transition between the different phases of a 3D-MOT task, showing some sort of a hand-off between attention and working memory.

\doublespacing

\section{Introduction} 
\label{sec:introduction}
Our ability to track multiple moving objects simultaneously in a dynamic environment enables us to perform everyday tasks such as driving, playing team sports, and walking in a crowded mall. In such tasks, it is required to manage internal representations of relevant information in order to predict future spatial positions of surrounding objects and optimize decision making accordingly. A professional athlete being able to make a perfect pass to a teammate in a high speed sport while avoiding players from the other team, is a great example of the brain's remarkable ability to track multiple objects both in space and time. 
In order to study this ability in a laboratory setting, researchers often use a variant of the multiple-object tracking (MOT) task developed by Pylyshyn \& Storm in 1988 (\cite{pylyshyn1988tracking}). In typical MOT tasks we find two categories of visual objects: targets (objects of interest) and distractors (objects to ignore) both sharing identical visual properties. In order to modulate task difficulty, parameters usually include the number of targets (\cite{pylyshyn1989role}), speed (\cite{chen2013resource, iordanescu2009demand}), and distance between objects (\cite{alvarez2007many, shim2008spatial}). 

Multiple-object tracking is an active area of research in humans but also in computer vision as we are observing an increasing demand for technology for automated tracking of vehicles and people in various contexts (\cite{milan2016mot16, dendorfer2020mot20, park2021multiple}). Humans' visual system has inspired current neural network architectures driving most of the artificial intelligence (AI) field as we know it today (\cite{rajalingham2018large}) and recent neural network architectures are trying to take advantage of higher brain mechanisms such as attention (\cite{luong2015effective}). Therefore, better understanding the underlying mechanisms of the brain's ability to track multiple objects in time and space could also benefit AI-related fields.

Despite the increasing literature on multiple-object tracking and evidence of effective training showing transfer on real tasks in real environments (\cite{faubert2012perceptual}), our understanding of the underlying neural mechanisms remains fuzzy. Several studies have shown that our capacity to keep individual items in working memory is limited to 3 or 4 (\cite{cowan2001magical, vogel2004neural}) and over the last few decades, several cognitive models have been brought forward trying to explain how individual items are being encoded and deciphering the intertwined roles of attention and working memory in such tasks. Older proposals suggested that early individuation of objects (up to 3-4) does not require attention mechanisms (\cite{trick1993enumeration}), however, recent research indicates otherwise suggesting that simultaneous indexing of items relies on attention mechanisms (\cite{cavanagh2011visual}) before involving subsequent mechanisms such as visual WM to encode the individuated objects in greater details. Another proposal include multifocal spatial attention, where attention can be split and work in parallel for multiple targets (\cite{cavanagh2005tracking}). On the other hand, Oksama and Hyöna suggested that rather than having a fixed-capacity parallel mechanism, the tracking performance would be better explained by a serial model where the maintenance of moving objects requires continuous serial refreshing of identity-location bindings (\cite{oksama2016position}). Moreover, a previous behavioural study also showed that one cognitive strategy is to process the targets as one illusory object by mentally creating connections between the targets to make, for example, a geometrical shape (\cite{yantis1992multielement}). 

The confounding aspects of attention versus working memory and location-based versus object-based tracking still remains fuzzy and recent EEG studies were performed trying to provide more insights by disentangling them. For example, Drew and colleagues tried to delineate the neural signatures of tracking spatial position and working memory during attentive tracking. They found that there was a unique contralateral negativity related to the process of monitoring target position during tracking which was absent when objects briefly stopped moving. These results suggest that the process of tracking target locations elicits an electrophysiological response that is distinct and dissociable from neural responses of the number of targets being attended (\cite{drew2011delineating}). Also, Merkel and colleagues looked at the spectral properties of the electrophysiological signal, mainly in the gamma range, during tracking to find a difference between location-based and object-based maintenance of visual information (\cite{merkel2022electrophysiological}). Their results suggest that object-based tracking is supported by enhanced encoding during the initial presentation of the targets and location-based tracking is characterized by the sustained maintenance of the individual targets during the entire tracking period, in that same processing neural network. In a previous study Merkel and colleagues also showed that neural networks involved in both tracking processes (object-based and location-based) are at least partly overlapping (\cite{merkel2015neural}). 

The experiment described in this manuscript is part of a larger study where we seek to develop a passive brain-computer interface composed of an EEG closed-loop system for a cognitively demanding task. We hypothesize that by looking at different neural correlates related to attention and working memory as well as other EEG signals in real-time while the user is performing such a task, we can identify different cognitive states and adapt the task in real-time to enhance the overall experience and outcome of the task. As part of the larger study, all subjects participated to three different tasks: MOT task, N-Back task, and flight simulator task. In this publication we share only our findings related to cognitive processes during the MOT task, and we don't cover the other tasks nor the BCI system which will all be discussed in depth in another publication.

Most MOT studies are conducted using a 2D-MOT task on a computer screen. However, we live in a 3D environment and different cognitive processes might be at play in a 3D environment. NeuroTracker\textsuperscript{TM} is a commercially available 3D-MOT task currently used by a multitude of users in many countries around the world as a perceptual-cognitive training and assessment tool. It has been used and studied in various fields such as sport (\cite{faubert2012perceptual, faubert2013professional, mangine2014visual, romeas20163d}), ESports (\cite{benoit2020neuropsychological}), education (\cite{tullo2018examining}), aviation (\cite{hoke2017perceptual}) and military (\cite{vartanian20163d}). 3D-MOT training has been demonstrated to enhance attention, working memory and visual information processing speed (\cite{parsons2016enhancing}). Given the wide adoption of the NeuroTracker\textsuperscript{TM} and existing literature showing effective transfer, we opted for a modified version of the NeuroTracker\textsuperscript{TM} for our study.

According to existing literature and the nature of the task, different neural correlates linked to working memory, attention, workload and visual processes should be at play during 3D-MOT. In the time domain, event-related potentials (ERPs) should be observed at the beginning of each phase of the task given the sudden change in visual information displayed. For lateralized trials, where the targets are displayed and moving only in one hemifield, we expect to see contralateral delay activity (CDA). The CDA is a sustained negativity over the hemisphere contralateral to the positions of the items to be remembered. As mentioned before, the CDA has been shown to be linked with the number of items held in WM (\cite{unsworth2015working, luria2016contralateral, roy2022cda}) and previous studies have shown a correlation between CDA amplitude and the number of targets during a 2D-MOT task (\cite{drew2008neural, drew2011delineating}). We hypothesise that during the indexing phase the participant would load the targets in working memory, hold the targets in working memory during tracking and then access them during recall to provide the answer. In the frequency domain, we are expecting different changes in frequency bands linked to working memory (e.g. Theta), attention (e.g. Alpha), workload (e.g. Beta), and visual processes (e.g. Gamma) during the different phases of the task.


\section{Materials and Methods} 
\label{sec:methods}

\subsection{Participants}
Twenty-four participants (thirteen females) aged between 21 and 41 years of age (M=29.3, SD=4.9) took part in this study. The participants were healthy university students of various ethnicity from different universities in Montreal. All participant self-reported normal or corrected-to-normal visual acuity and passed the Randot Stereotest for stereo vision. The  study  was reviewed and approved  by  the  Université  de  Montréal  ethics  committee  for  health  research (Comité d’éthique de la recherche en santé; CERES \#2018-334). Recommended ethics procedures and guidelines were followed, and informed consent was obtained from all participants. The three hour long session included the 3D-MOT task discussed here but also included the recording of a N-Back task and both tasks were part of a larger research project where the subjects also participated in a second session, on a different day, for a flight simulator task. All subjects received a monetary compensation for their participation to the two sessions covering also their transportation to the different facilities. The other components of the larger research project (i.e. N-Back and flight simulator) are not discussed in this manuscript and will be published separately.

\subsection{Task} 
\begin{figure}[!htb]
    \centering
    \includegraphics[width=\textwidth]{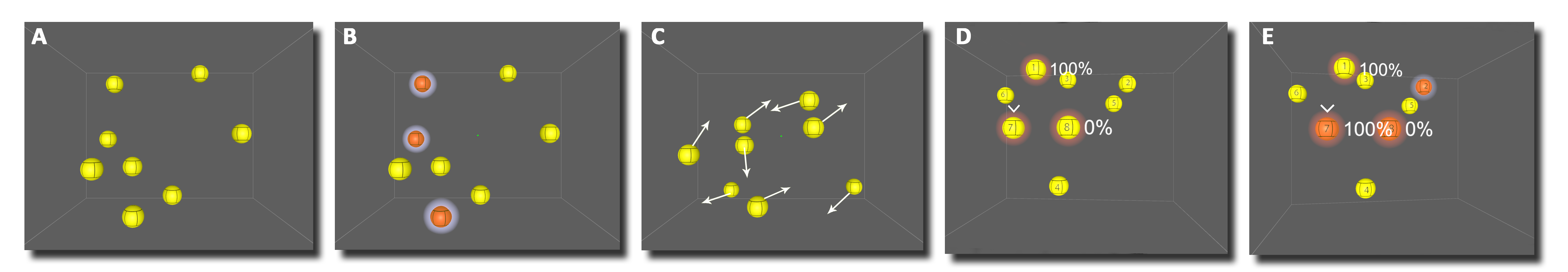}
    \caption{3D-MOT Task Sequence. (A) All spheres appear on screen. (B) Targets are highlighted in red for 2 seconds. (3) All the spheres are moving for 8 seconds. (D) Participant must identify the targets and provide a confidence level. (E) Feedback is provided to the participant showing the correct answers.}
    \label{3D-MOT_Sequence}
\end{figure}

Figure \ref{3D-MOT_Sequence} shows the five different phases of the 3D-MOT task developed with the Unity engine. (A) \textit{presentation phase} where 8 yellow spheres are shown in a 3D volume space for 2 seconds, (B) \textit{indexing phase} where one, two or three spheres (targets) change colour (to red) and are highlighted (halo) for 2 seconds, (C) \textit{tracking (or movement) phase} where the targets indexed in phase 2 return to their original colour (yellow) and 1 second later start moving for 8 seconds crisscrossing and bouncing off of each other and the virtual 3D volume cube walls, (D) \textit{recall phase} where the spheres stop moving and the observer is prompted to identify the spheres originally indexed in phase 2. Each sphere is labelled with a number between 1 to 8. After identifying the targets, the observer is asked to provide a confidence level for each answer (either 0\%, 25\%, 75\% or 100\% confident). And finally, (E) \textit{feedback phase} where the correct targets are clearly identified on the screen. The whole trial takes around 15s (2s + 2s + 9s + ~[1-4]s) depending on how long the participant takes to provide the answers. A video of the task is available online on the repository provided.

The participant was seated 1.75m and centered from the 65" 3D TV screen (Panasonic TC-P65VT60) wearing the EEG cap as well as active shutter 3D glasses (Panasonic TY-ER3D5MA). The 3D-MOT virtual cube was 35 degrees of visual angle (dva) in size  and the spheres were 2.5 dva. The 3D glasses were carefully slid under the EEG cap in a way to minimally disrupt the EEG signal. Additional gel was added to some temporal electrodes for some participants to compensate for the gap created by the 3D glass legs. The participants had a keyboard on their lap to provide their confidence level after giving the answers orally to the instructor, seated 2m behind the participant with a keyboard. 

Each participant started with a training block of 20 trials, without the EEG equipment, to familiarize themselves with the task and for us to obtain an individualised speed threshold. They were instructed to keep their eyes on the green fixation dot in the middle of the screen and to let their covert attention track the targets. An adaptive staircase algorithm modulated the speed of each new trial based on previous performances to find the individual speed where the participant gets a performance of 50\% (i.e. they get half the trials right and half wrong) (\cite{faubert2012perceptual}). To get a \textit{good} trial the participant needs to identify correctly all the targets. For the training block, only the condition with 3 targets was used. After the 20 trial training block, the resulting speed was used for the remaining of the experiment.

The experiment consisted of 4 blocks of 21 trials with 2 conditions: side and set size. The speed was kept constant, based on the speed obtained at the end of the training block. In total, 30 trials were presented in the left hemifield, 10 for each set size (1,2,3), 30 trials were presented in the right hemifield, 10 for each set size (1,2,3), and 24 trials were not lateralized and the targets could freely cross from left to right and vice-versa, 8 for each set size (1,2,3).  It is important to note that for lateralized trials only the targets stayed on one specific hemifield while the other distractors where moving freely with no restriction. Once the targets stopped moving, a number between 1 and 8 appeared on each of the spheres and the participants had to provide their answer by saying the number of the target(s) out loud for the instructor to enter the answers. After the answers were entered, a visual cue on top of each of the selected spheres appeared for the participant to provide a confidence level for each target. The participant used the arrow on the keyboard to provide the confidence level. Up arrow means 100\% confident (the instruction provided to the participant: "I was able to track it."). Down arrow means 0\% ("I lost it, it's a random guess."). Right arrow means 75\% confident ("I'm somewhat confident. I think I tracked it but I might have switch target during an overlap/occlusion."). Left arrow means 25\% confident ("It's mostly a guess. I got confused."). The rational behind this discrete scoring of confidence was to clearly dissociate between a random guess and a fully confident answer (0\% vs 100\%). For the in-between, we decided to avoid having only one additional option as we felt like most participants might default to that option whether they are quite confident but do have a small doubt or if they are mostly guessing it, making this option hard to use for further analysis. Adding more granularity than four options for the confidence level would have only added a cognitive load with no additional benefit. Therefore the compromise with four discrete options was chosen. For the analysis presented in this manuscript, we regrouped 75\% and 100\% as being a confident answer and therefore a \textit{good trial} is a trial where the participant identified all the correct targets and indicated a confidence level of 75\% or 100\%, otherwise it is labeled as a bad trial.

\subsection{EEG Acquisition}
The electroencephalogram (EEG), electrocardiogram (ECG), and electrooculogram (EOG) were recorded using the Biosemi ActiveTwo system (Biosemi, Amsterdam, Netherlands) with 71 Ag–Ag/Cl electrodes positioned at 64 standard International 10/20 System sites (EEG), left and right mastoids for offline EEG re-reference, 1cm lateral to the external canthi for horizontal EOG (HEOG), left and right ribs plus right collarbone for ECG. The HEOG was used for eye movements to confirm that the participant was tracking the targets with covert and not overt attention (i.e. not moving their eyes). Electrophysiological signals were digitized at 2048Hz.

\subsection{EEG Analysis}
EEG offline analysis was performed in MNE-Python (\cite{gramfort2013meg}) an open-source Python package for neurophysiological data. Both our EEG data and code are available online. Here we will breakdown the analysis in two parts, first in the time domain and second, in the frequency domain to look at the oscillation variations over time and power spectrum across conditions. Four conditions were examined across different brain regions: side (left, right), set size (1, 2, 3), performance (good, bad), 3D-MOT phase (indexing, tracking, recall). Note that because of low amount of trials with bad performances, we looked at the data internally but do not draw any conclusion in this paper about performance. It is important to note that after looking at our results we realized that we unfortunately had a 150ms offset in our Unity code between the trigger (i.e. EEG event) and the color change (from yellow to red and red back to yellow) appearing on screen. The offset is fixed and due to a Unity fixed update delay we forgot to remove for the color change. This offset does not apply for when the movement stops and ends. What it means is that the events at t=0s, and at t=2s, are in fact visible on screen only at t=0.15s and t=2.15s. \\

\textbf{Preprocessing.} First, the EEG channels were re-referenced to the left and right mastoids. Second, independent component analysis (ICA) was used to remove eye blinks and eye movement artifacts. Third, the EEG data was epoched in [-1, 15]s windows where t=0s represents the stimuli/trigger of the spheres being highlighted in red. Fourth, AutoReject (\cite{jas2017autoreject}) was used to automatically remove bad trials and correct bad channels. After looking at the clean EEG data and separately analyzing EOG channels for eye movements during tracking to see if people were really tracking with covert attention instead of overt attention, 4 subjects were removed and 20 subjects remained for the analysis. 

\textbf{Time domain.} First, the non-lateralized occipital grand average signal was obtained over O1, O2 and Oz to confirm visual ERPs. Since the 3D-MOT is a visual task with sudden visual changes between the different phases, a visual/occipital ERP should be observed accordingly. Second, before looking at the CDA we seeked to confirm that the brain activity was lateralized for left vs right trials where targets were presented and moved only in one hemifield. For the lateralized activity, we looked at different clusters of electrodes, namely \textit{frontal}, \textit{central}, \textit{parietal}, \textit{temporal} and \textit{occipital}. For readability the electrodes used in each cluster are listed in Table \ref{table_clusters-electrodes} in supplementary material. For the frontal cluster, all channels with the letter 'F' were included. For the central cluster, all the electrodes with the letter 'C' were included. For the parietal cluster, all the channels with the letter 'P' were included. For the temporal cluster, all the channels with the letter 'T' were included. For the occipital cluster, all the channels with the letter 'O' were included. For lateralized activity of both left and right trials, we averaged the left channels (i.e. channels with odd numbers) and subtracted the average of the right channels resulting in the difference between the two hemispheres. The rationale behind this analysis is to confirm that indeed we see more activity in one hemisphere than the other for lateralized trials. The gross activity observed here should encompass specific neural signatures such as ERPs and CDA, which are then analyzed in greater detail.

Third, we looked at the effect of the number of targets on the CDA by averaging of the channels contralateral to the hemifield where the targets were presented minus the average of the ipsilateral channels. The midline channels (ending with the letter \textit{z}) were not included in the CDA nor the lateralized analysis. The CDA, as defined in the literature, should be strongest in the parietal region. However, we also calculated the same activity (i.e. contra minus ipsi) for the different clusters mentioned above. For both lateralized activity and CDA we used only the trials with a good performance (i.e. the participant identified all the targets correctly with high level of confidence) for a total of N=982 trials, an average of 49 trials per subject. A three-way repeated measures analysis of variance (ANOVA) was performed with the number of targets (1, 2, and 3), phases (id, tacking, and recall), and clusters (frontal, central, parietal, temporal, and occipital) as independent variables and the mean CDA amplitude over a 1s time window as the dependent variable. For the identification (id) phase, the time window was from 0.5s to 1.5s, for the tracking we selected the 5s to 6s and for the recall, we used 11.5s to 12.5s.

\textbf{Frequency domain.} To obtain a more detailed representation of neural oscillation changes over time, Event-Related Spectrum Perturbation (ERSP) graphs were used. The time-frequency decomposition was computed using Morlet wavelets for frequencies between 1 and 50Hz with varying cycles of half the frequency. The ERSP maps were then obtained by getting the log ratio of the power relative to the baseline power. For the time-frequency analysis, the baseline was selected as -1s to 0s prior to the targets being colored in red (t=0.15s). Instead of using raw power, the log ratio has the advantage of normalizing the power across participants. To investigate the potential role of different brain regions we used the midline channels from frontal to occipital to give a representation in space of the time-frequency activity. For this analysis we included only the trials with a good performance. With the same time-frequency decomposition, we analyzed a 1s window of average power as a dependent variable using a repeated measure three-way ANOVA, with Phase (Identification [0.3, 1.3]s, Tracking [5, 6]s, and Recall [11.5, 12.5]s), Set Size (1, 2, and 3), and Frequency Band (delta [1-3]Hz, theta [4-7]Hz, alpha [8-12]Hz, beta [13-30]Hz, gamma [31-50]Hz) as independent variables. The time windows across the phases were selected to capture mostly top-down cognitive processes.


\section{Results} 
\label{sec:results}

\subsection{Time Domain}

\begin{figure}[!htb]
    \centering
    \includegraphics[width=0.9\textwidth]{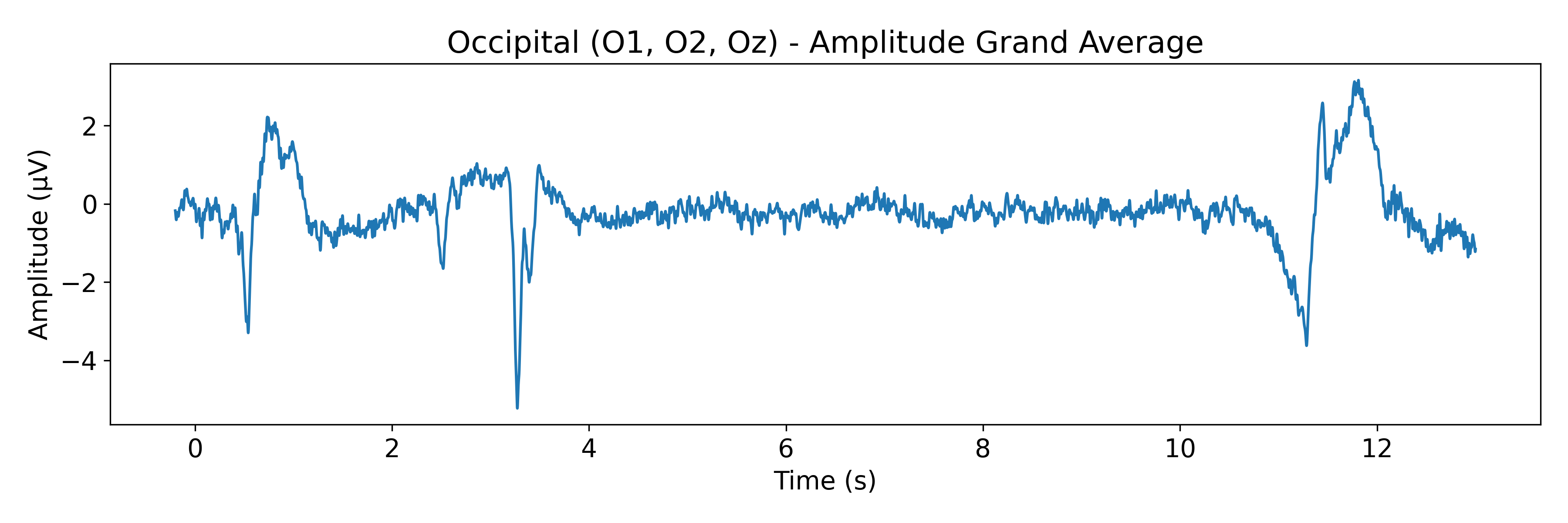}xw
    \caption{Non-lateralized occipital ERPs (channels: O1, O2, Oz). Grand average over all participants for all conditions and performances.}
    \label{Occipital-ERP}
\end{figure}

Non-lateralized occipital activity in O1, O2 and Oz (see Figure \ref{Occipital-ERP}) confirmed the visual ERPs when drastic visual changes occured, such as the target spheres changing color from yellow to red at t=0.15s, then reverting back to yellow at t=2.15s, start moving at t=3s and stop moving at t=11s. Noteworthy, the occipital ERP of the initial color change from yellow to red is significantly stronger than when the targets revert back to yellow (t-test on 200ms window; p $<$ 0.001) and the strongest ERP occurs when the targets start moving. On the other hand, the ERP elicited when spheres stop moving isn't as sharp as the other ones, most likely due to the time it takes for individuals to realize that the spheres have indeed stopped moving. Moreover, we can also see a sustained cognitive activity following the ERPs at t=0.15s and t=11s, respectively during identification of targets and recall.

\begin{figure}[!htb]
    \centering
    \includegraphics[width=0.9\textwidth]{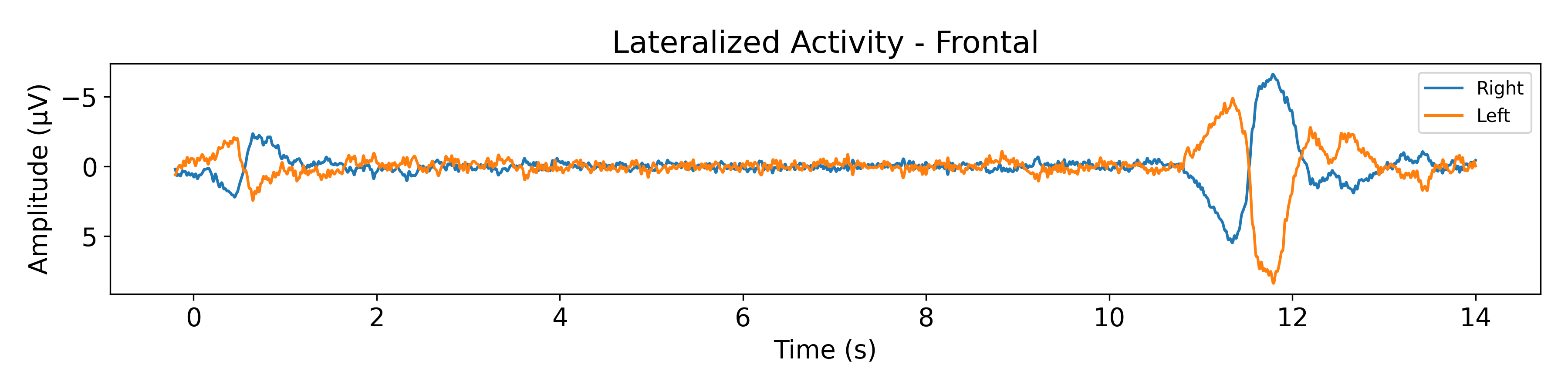}
    \caption{Lateralized activity in the frontal region for whole sequence. Left vs right trials.}
    \label{Lateralized-Activity-FS}
\end{figure}

\begin{figure}[!htb]
    \centering
    \includegraphics[width=0.45\textwidth]{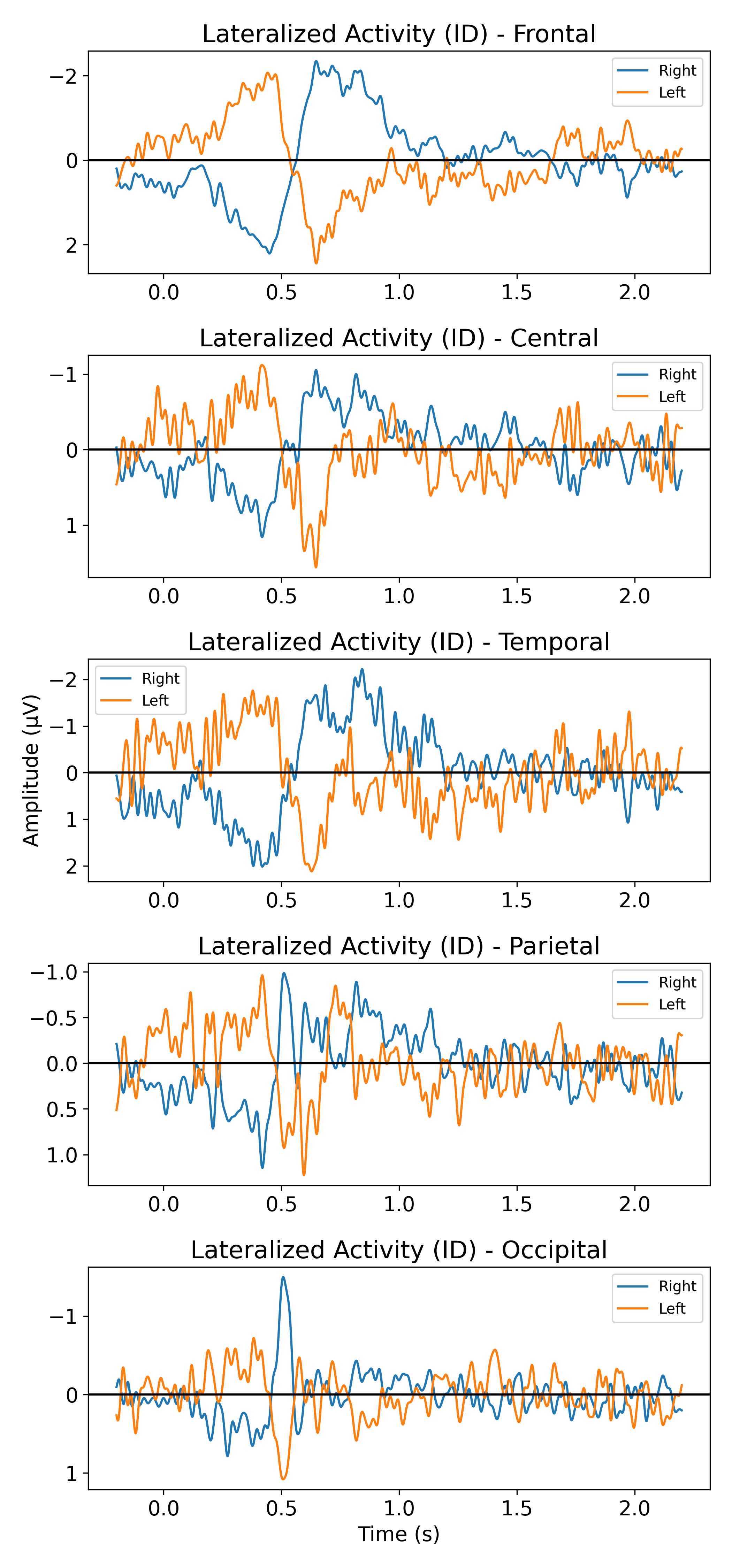}\hfill
    \includegraphics[width=0.45\textwidth]{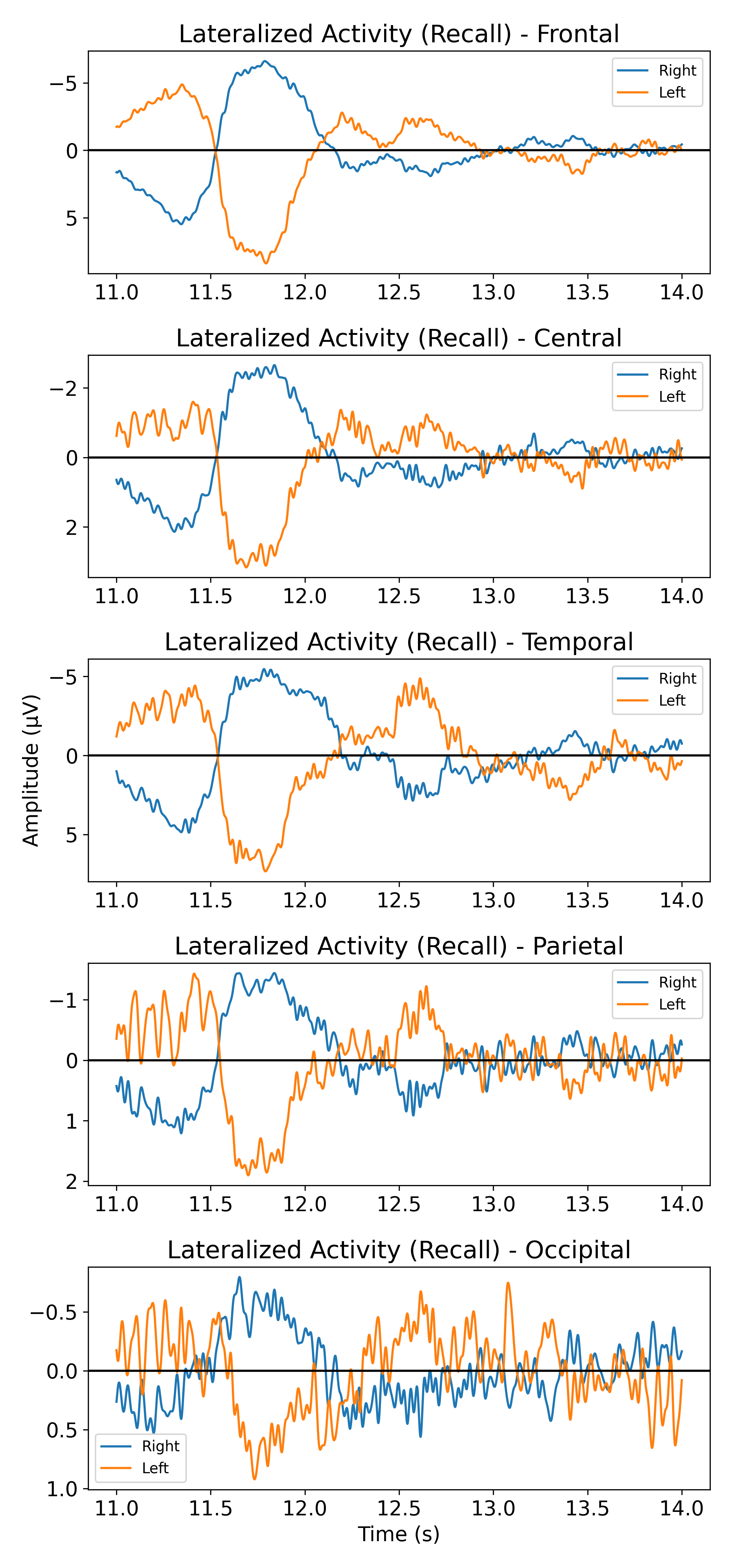}
    \caption{Lateralized activity from the different brain regions for identification (first column) and recall (second column) phases. Left vs right trials.}
    \label{Lateralized-Activity-ID-Recall}
\end{figure}

As expected, we obtained a clear lateralization of activity in left vs right trials during both identification and recall phases as shown on Figure \ref{Lateralized-Activity-FS}. However, we did not observe lateralized activity during tracking despite the targets staying in only one hemifield for the full duration of the trial. Zooming in on the identification and recall phases, we observe that the same brain regions (clusters) are activated the strongest, namely the frontal and temporal regions, during both identification and recall (see Figure \ref{Lateralized-Activity-ID-Recall}). This could be indicative of similar cognitive functions during identification and recall phases. Note that automated scales were used to preserve a clear shape of the signal, however the amplitude, or strength of the signal, varies across clusters.

\begin{figure}[!htb]
    \centering
    \includegraphics[width=0.9\textwidth]{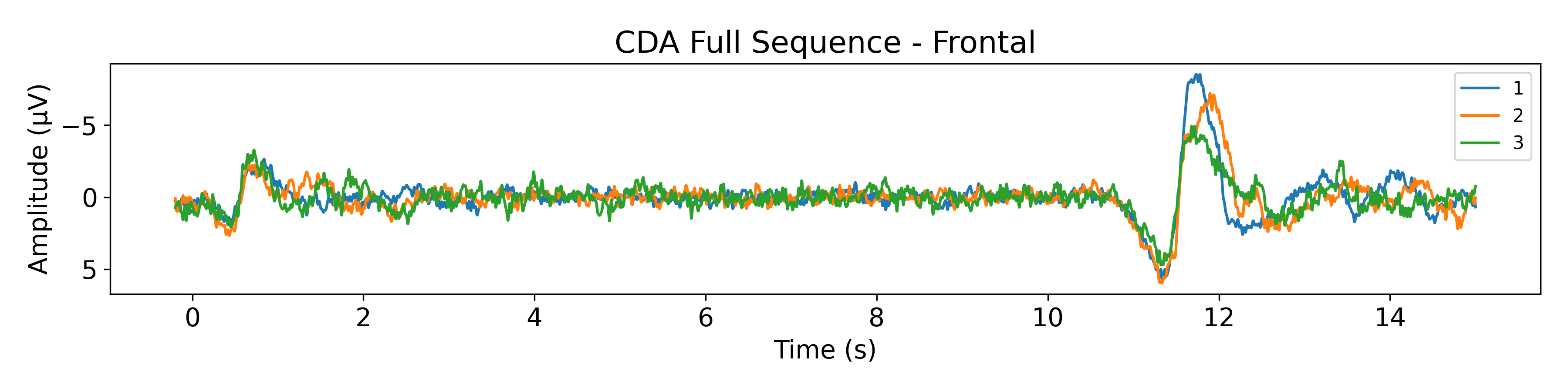}\hfill
    \caption{CDA in the frontal region for the full sequence. Trials with 1, 2 and 3 targets.}
    \label{CDA-FullSequence-Frontal}
\end{figure}

\begin{figure}[!htb]
    \centering
    \includegraphics[width=0.45\textwidth]{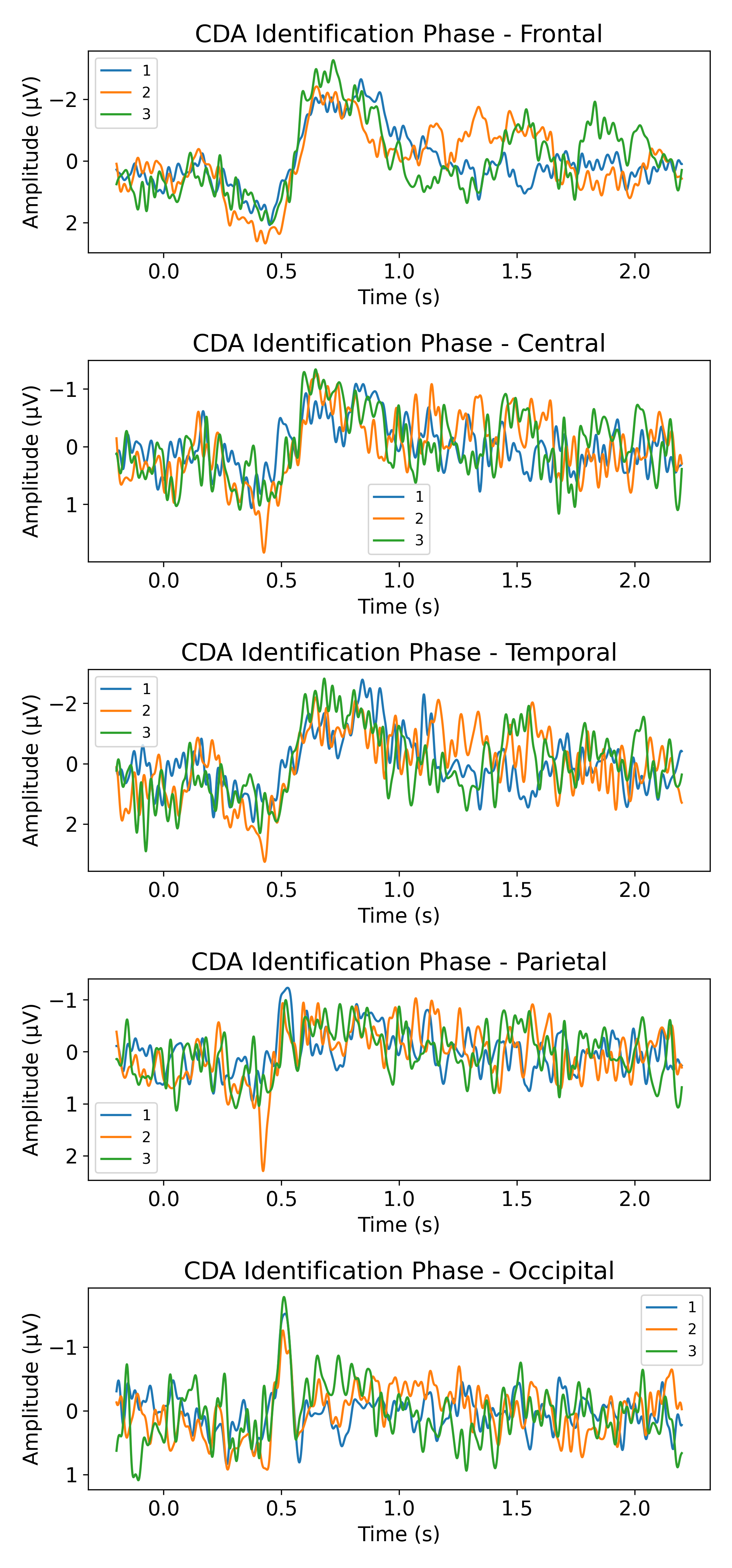}
    \includegraphics[width=0.45\textwidth]{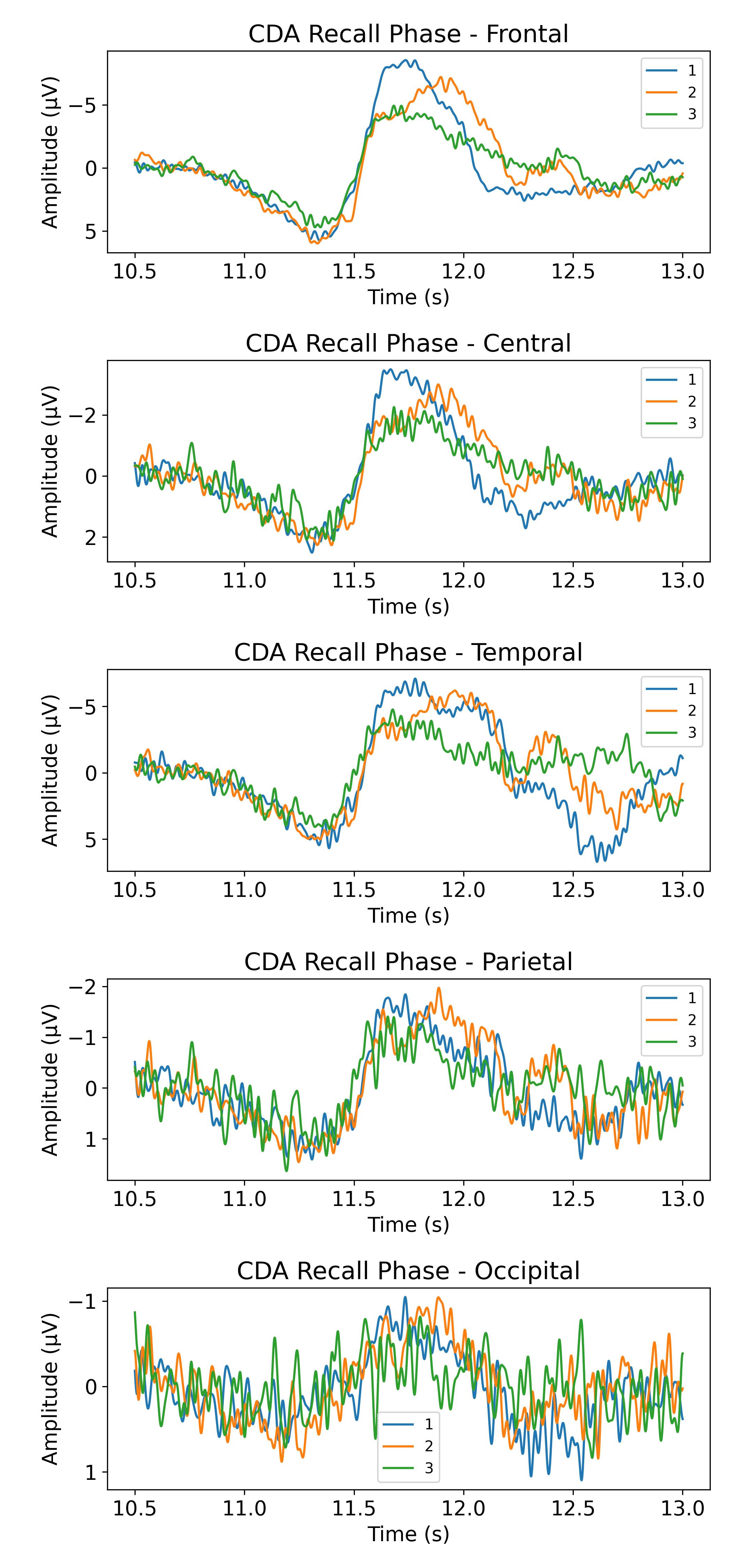}
    \caption{Lateralized activity from the different brain regions for identification (first column) and recall (second column) phases. Trials with 1, 2 and 3 targets.}
    \label{CDA-ID-Recall}
\end{figure}

Figure \ref{CDA-FullSequence-Frontal} shows the CDA amplitude for the frontal cluster for the whole duration of the trial and on Figure \ref{CDA-ID-Recall} we zoom in on the identification and recall phases across EEG channel clusters. The contralateral delay activity observed here has a similar shape as what we see in the literature, peaking around 500-600ms post-stimuli and slowly decaying for another $\sim$500ms but we expected the strongest CDA in posterior parietal regions as opposed to frontal regions (\cite{vogel2004neural, unsworth2015working}). The initial three-way repeated measures ANOVA with the number of targets (1, 2, and 3), phases (id, tacking, and recall), and clusters (frontal, central, parietal, temporal, occipital) as independent variables and the mean CDA amplitude as the dependent variable, revealed a significant effect for the number of targets (F\textsubscript{(2,34)} = 3.59, p = .039), a strong significant effect for the phase (F\textsubscript{(2,34)} = 20.17, p $<$ .0001), a strong significant effect for the cluster (F\textsubscript{(4,68)} = 32.85, p $<$ .0001), a near significant effect for the interaction between the targets and phases (F\textsubscript{(4,68)} = 2.46, p $=$ .053), a non significant effect for the interaction between targets and clusters (F\textsubscript{(8,136)} = 1.56, p = .14), a strong significant effect for the interaction between phases and clusters (F\textsubscript{(8,136)} = 18.3, p $<$ .0001), and a trending but non significant effect for the interaction between the three (F\textsubscript{(16,272)} = 1.6, p = .067). Post hoc one-way ANOVAs looking only at the number of the targets, independently for each phase and cluster, revealed no significant effect on the mean CDA amplitude during the identification phase nor during tracking, in any of the clusters. During recall, however, the effect was significant in the parietal cluster (F\textsubscript{(2,34)} = 8.15, p $<$ .005), the temporal cluster (F\textsubscript{(2,34)} = 7.76, p $<$ .005), the central cluster (F\textsubscript{(2,34)} = 5.59, p $<$ .01), the occipital cluster (F\textsubscript{(2,34)} = 6.93, p $<$ .005) and trending but non significant in the frontal cluster (F\textsubscript{(2,34)} = 2.6, p $=$ .088). Note that the post hoc ANOVA values are provided as is and were not corrected for multiple comparisons. A total of 15 post hoc ANOVAs were performed (5 clusters x 3 phases). After obtaining and writing the results, We also performed the same analysis with a 0.5s window to see if it would change the results but it yielded similar results.

\subsection{Frequency Domain}
\begin{figure}[!htb]
    \centering
    \includegraphics[width=0.5\textwidth]{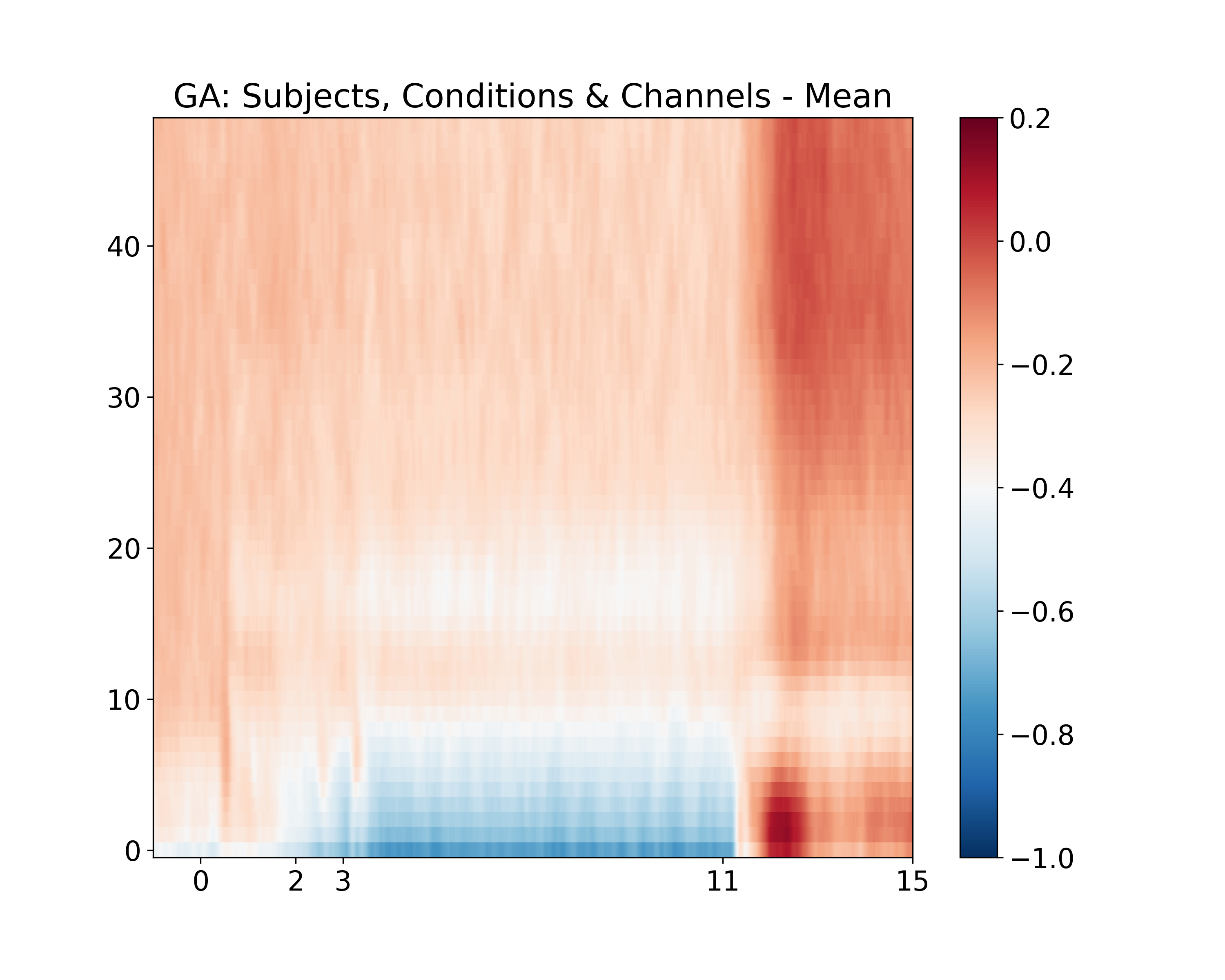}
    \caption{ERSP: Grand Average (GA) across all subjects, all conditions and all channels. The color represents the log ratio of the power at each instant with average power of the baseline [-1,0]s for that same frequency, averaged across subjects.}
    \label{ERSP-GrandAverage}
\end{figure}

The ERSP for the whole trial is shown on Figure \ref{ERSP-GrandAverage} which includes 1 second prior to the targets being identified in red and up to 15s post identification, with frequencies ranging from 1 to 50Hz. It represents the grand average across all participants, all conditions, and all channels. Before breaking down the time-frequency analysis to look at specific elements, we can easily distinguish the different phases of the MOT task with drastic changes in neural oscillations across these phases. It is important to note the range of the color scale of the ERSP maps presented in it section as in EEG studies blue usually means a desynchronization (i.e. negative value) and red a synchronization (i.e. positive value). Here, we purposefully used a none symmetrical color scale to accentuate the differences observed given that most of the values are negative.

On the spectral map (Figure \ref{ERSP-GrandAverage}) we can identify, as expected, a perturbation near t=0.15s when the spheres turn red, creating a narrow perturbation corresponding to the ERP in the time domain related to the sudden visual change. Also visible are perturbations from the ERP near t=2.15s when they turn back yellow and from the ERP near t=3s when they suddenly start moving. During tracking, we observe a strong inhibition of lower frequencies until alpha band ($\sim$8Hz) where the inhibition isn't as strong, then the inhibition seems to continue in the $\sim$14-18Hz range after which there is a change in inhibition intensity for high beta and gamma. More strikingly, we observe a clear cognitive switch as soon as the spheres stop moving and the participant is asked to answer (at t=11s). Suddenly, the lower frequencies are re-activated and alpha slightly reduced. After t=13s, the values can't really be interpreted given the variability in speed to answer from trial to trial, across participants and the number of targets.

\begin{figure}[!htb]
    \centering
    \includegraphics[width=0.8\textwidth]{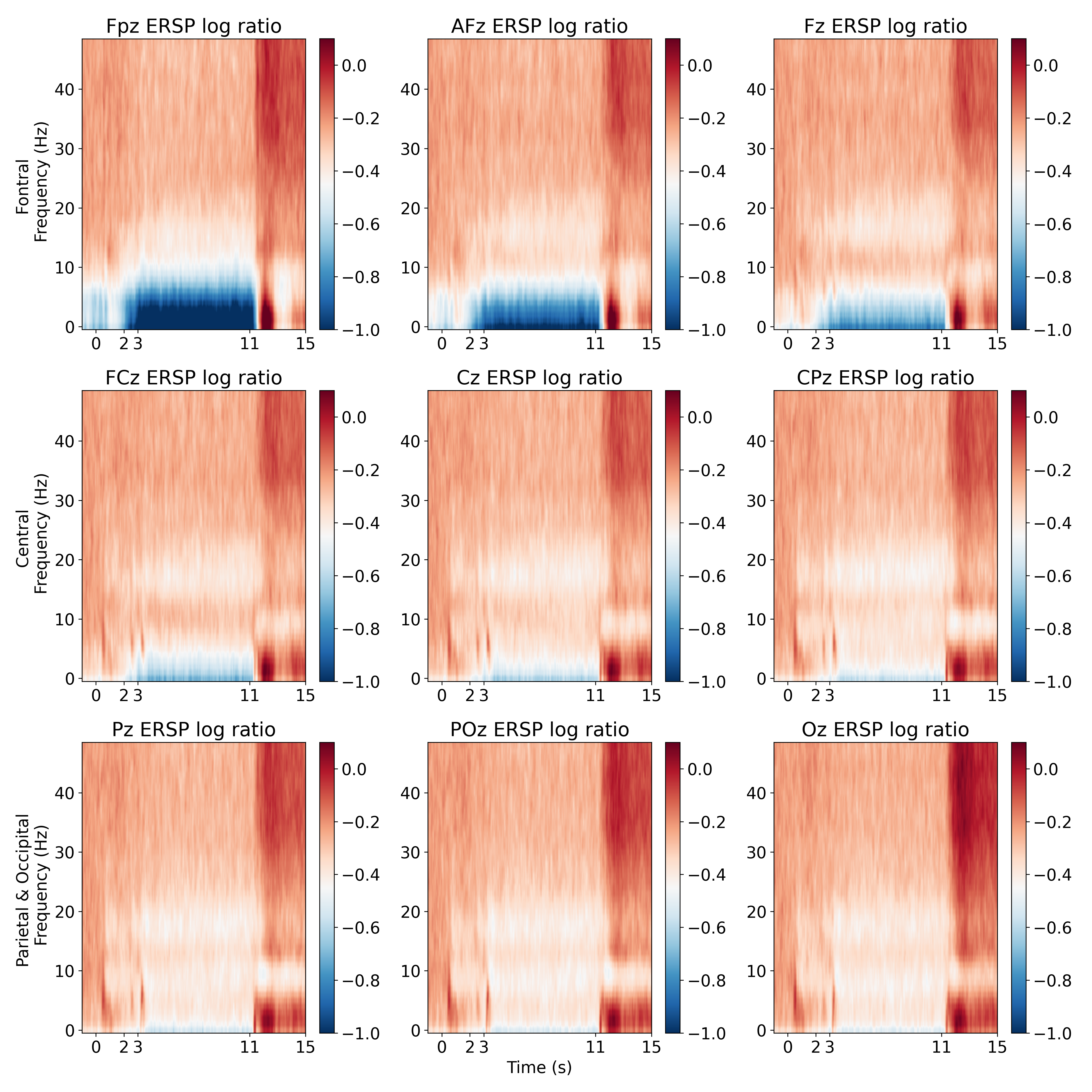}
    \caption{ERSP: Midline electrodes, frontal to occipital. The color represents the log ratio of the power at each instant with average power of the baseline [-1,0]s for that same frequency, averaged across subjects.}
    \label{R2a-ERSP-MidLine}
\end{figure}

Looking at the spatial distribution over the midline channels on Figure \ref{R2a-ERSP-MidLine}, we observe that during tracking (t=[3-11]s), there is a strong delta and theta inhibition in the frontal regions. During recall (t=[11-13]s), we observe a sudden re-activation of these lower frequencies in the delta and theta bands from frontal to occipital regions. During recall, we also observe a strong activation of high-beta to low-gamma in the occipital region.

\begin{figure}[!htb]
    \centering
    \includegraphics[width=0.8\textwidth]{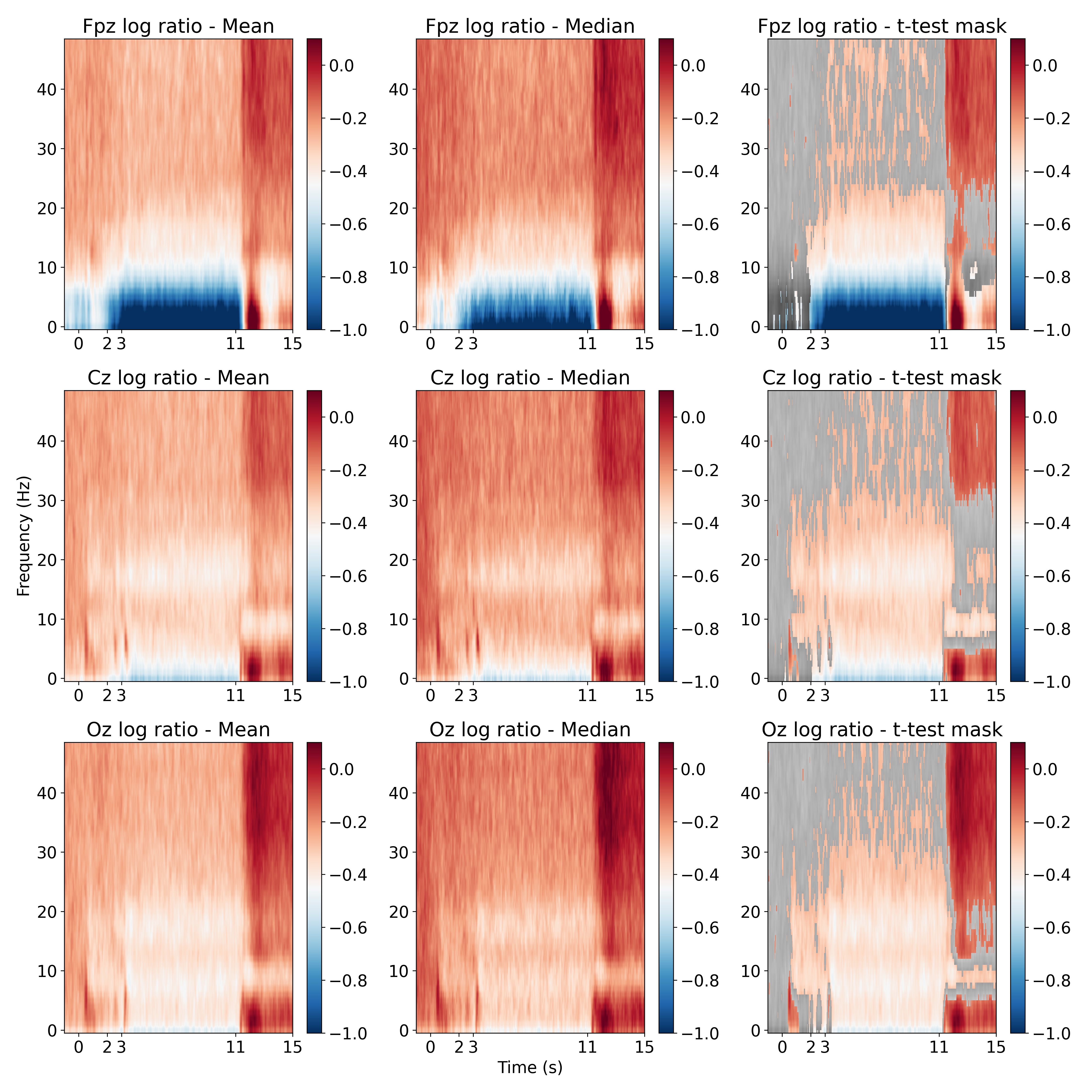}
    \caption{ERSP: Fpz, Cz, and Oz electrodes. The color represents the log ratio of the power at each instant with average power of the baseline [-1,0]s for that same frequency. The first column is the mean across subjects. The second column is the median across subjects. The third column is the mean across subjects with a gray mask where the p-value of a t-test $>$=.05 (i.e. gray means not significantly different than baseline).}
    \label{R2b-ERSP-MidLine}
\end{figure}

To better understand the effect across participants, both the mean and the median values were calculated for each time-frequency points and what we observe on Figure \ref{R2a-ERSP-MidLine} is that the mean is not being distorted by outliers and that the distribution across trials and participants is somewhat symmetrical. A paired t-test was done for each time-frequency point comparing with the mean power of the baseline and the time-frequency points with p $>$= .05 were grayed-out on Figure \ref{R2b-ERSP-MidLine}. For readability we show the statistical analysis of only three channels (Fpz, Cz, Oz) as the six others were showing similar overall trends. We did not perform any multiple comparison correction, so given that we have performed tens of thousands of comparisons (i.e. for each time-frequency point independently) we are exposed to familywise type 1 error, meaning that the graphs presented here with the statistical masks are most likely showing more significant effects then there really are. Performing multiple comparison corrections like false discovery rate (FDR) on such a high amount of comparisons would obviously result in the opposite and hide all effects (type 2 error). Different approaches have been suggested to deal with ERSP significance such as cluster-level statistical permutation tests, however, given the mean and median of the log ratio with the baseline being both similar and biologically sound in time and frequency, correcting for the comparisons would most likely not invalidate the general trends we are observing here. Also noteworthy, as we can see on the ERSP graphs, the edge between alpha and theta bands around 8Hz and the edge between alpha and beta bands around 12Hz are blurry and comes out as non-statistically significant (vs baseline). This is most likely due to the individual differences in frequency bands (\cite{haegens2014inter}).

\begin{figure}[!htb]
    \centering
    \includegraphics[width=0.8\textwidth]{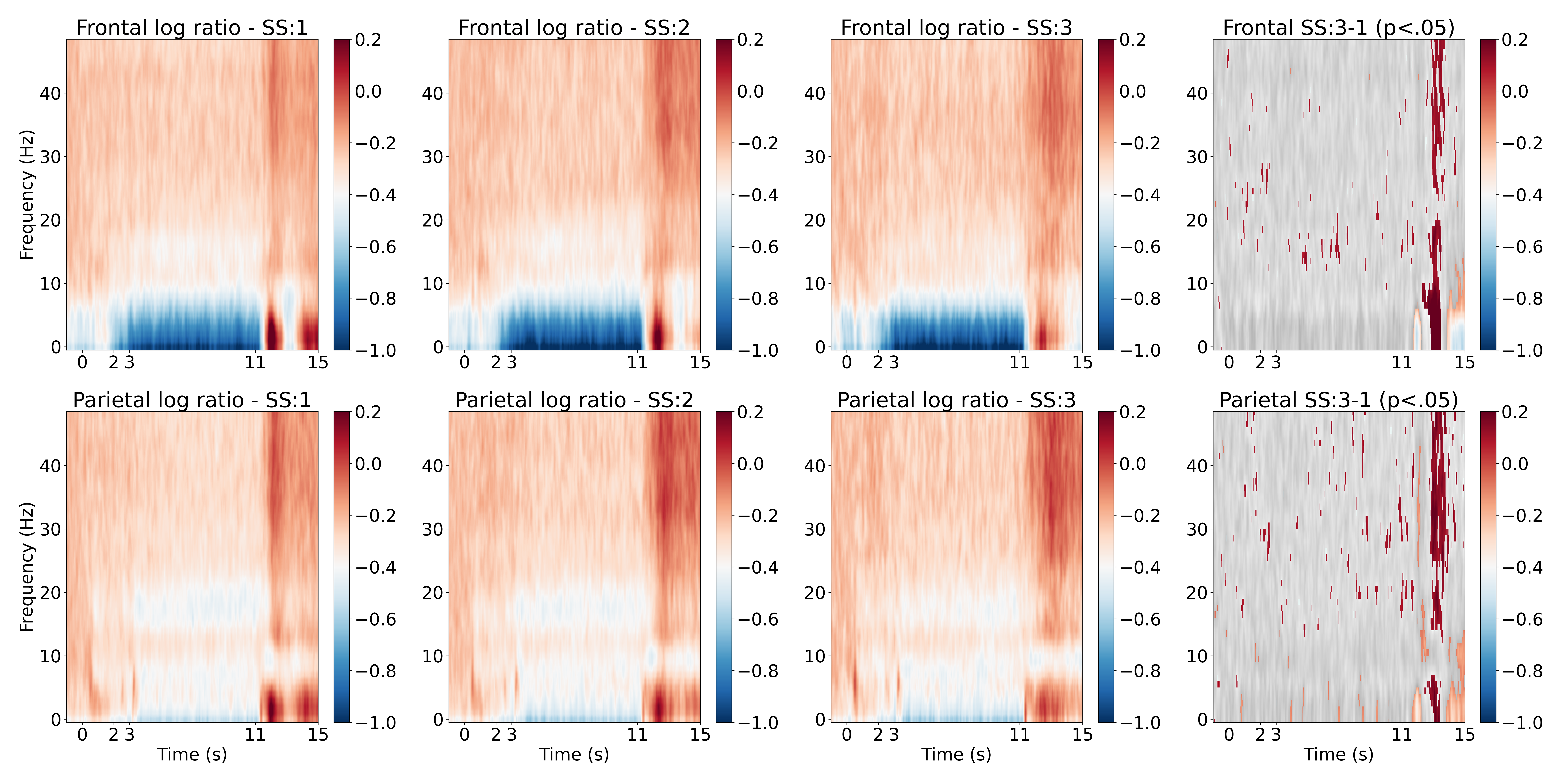}
    \caption{ERSP: Frontal and parietal regions for 1, 2, and 3 targets. The color represents the log ratio of the power at each instant with average power of the baseline [-1,0]s for that same frequency. The first column is the mean across subjects. The last column is the ERSP with 3 targets minus the ERSP with 1 target with a gray mask where the p-value of a t-test between the two $>$=.05 (i.e. gray means no significant difference between set size of 1 target vs 3 targets.)}
    \label{R3-ERSP-SetSizes}
\end{figure}

The analysis for the effect of the number of targets showed a significant difference only during recall. ERSP values for 1, 2 and 3 targets were obtained and paired t-tests were performed for condition 1 vs 3 targets on the whole trial for each time-frequency point. The statistical mask shown on the right on Figure \ref{R3-ERSP-SetSizes} grays out the points for which the p-value was higher or equal to 0.05, and therefore, not significantly different from 1 to 3 targets. The mask is plotted on top of the resulting ERSP map obtained by subtracting the ERSP values of 1 target to the ERSP values of 3 targets to highlight the difference in power between the two. The statistical tests for 1 vs 2 and 2 vs 3 targets were not performed because we assume some sort of linearity between 1, 2 and 3 targets and therefore their effect would be somewhere between what we observe for 1 and 3 targets. For readability only the frontal and parietal clusters are displayed. Frontal because that's where we observed the strongest CDA in the time domain and parietal because that's where we initially expected the strongest difference. 

Finally, as expected given the results from the ERSP maps showed and explained above, the repeated measure three-way ANOVA on raw power revealed a significant effect for Phase (F\textsubscript{(2,36)} = 12.13, p $<$ .0001), for Frequency Band (F\textsubscript{(4,72)} = 50.4, p $<$ .0001), and smaller yet significant effect for Set Size (F\textsubscript{(2,36)} = 4.6, p = .016). The interaction between Phase and Frequency Band was significant (F\textsubscript{(8,144)} = 11.58, p $<$ .0001), the interaction between Phase and Set Size was significant (F\textsubscript{(4,72)} = 6.36, p $<$ .0005), the interaction between Frequency Band and Set Size was significant (F\textsubscript{(8,144)} = 5.24, p $<$ .0001) and finally, the interaction between Phase, Set Size and Frequency Band was also significant (F\textsubscript{(16,288)} = 6.71, p $<$ .0005). The statistical analysis was performed on both the raw power (results mentioned above) and the power ratio (ERSP maps on figures). We reported the results from the raw power as most studies use the raw power directly. Using the power ratio instead of raw power provides a more normalized value across subjects, and we obtained similar results. The differences include the Set Size effect that came out slightly over the p = .05 significance threshold, with (F\textsubscript{(2,38)} = 2.92, p = .066), the interaction between Set Size and Phase was trending towards significance (F\textsubscript{(4,76)} = 2.46, p = .0521) and the interaction between Set Size, Phase and Frequency Band was significant (F\textsubscript{(16,304)} = 2.88, p $<$ .0005).


\section{Discussion} 
\label{sec:discussion}

Several studies have been published on EEG activity during MOT tasks (e.g. \cite{drew2008neural, bland2020gamma, merkel2022electrophysiological}), usually focusing on either the indexing or tracking phase but very little, if at all, on the recall phase. Here, we took a more holistic view to look at electrophysiological changes over the whole sequence (i.e. trial).

\textbf{CDA}. During the identification and tracking phases, we did observe lateralized activity as expected. However, the highest amplitude were seen in frontal and temporal regions and not in the parietal region as expected from the recent CDA literature. Also, the amplitude in the parietal region did not increase with the number of targets during identification and tracking which is in contradiction with the recent CDA literature (\cite{vogel2004neural, drew2008neural, unsworth2015working}) and our own CDA review paper (\cite{roy2022cda}). Moreover, during recall, we actually observed the opposite effect where the amplitude in certain regions was actually higher with less targets and the activity lasted longer in time with more targets. Given the clear lateralized ERPs and activity observed in the results (e.g. Figure \ref{Lateralized-Activity-FS}) the design of the task clearly had laterialized targets. Therefore, to explain our results being different from what we expected based on the literature, we rule out a bad task design and point out four ways in which our 3D-MOT task is different than most VWM tasks used to study CDA. First, as opposed to change detection tasks where the participant has to remember all the items at once, here the participants could have recalled the items one by one, sustaining a longer CDA over time. Second, most VWM tasks in CDA studies have very short trials of about 1 to 2 seconds long. Third, in most CDA related tasks, the objects are temporarily removed from the screen forcing the participant to have an internal representation of the objects. Here the objects are always visible for the participant and perhaps engage a different cognitive strategy. Forth, most tasks are done in 2D without stereoscopy. These confounding differences could make the participant use a different cognitive strategy relying more on attention mechanisms than working memory during tracking. 

In order to better understand the results we obtained, two things have to be disentangled. First, is WM required for the 3D-MOT task and if so, are the targets held in memory for the whole duration of tracking or only held in memory during indexing and/or recall? Second, is the lateralized activity observed here the same as the CDA referred to in the literature and believed to play a role in WM, or is it a different, yet lateralized, activity? If the answer is yes to both, as we initially hypothesized, we should have observed a CDA during tracking and the CDA amplitude should have varied in amplitude based on the number of targets. Our results diverge from \cite{drew2008neural} where they showed a clear CDA with different amplitudes during tracking in a different and shorter 2D-MOT task. As we can see in our results, the activity is clearly lateralized and has a CDA-like shape, in both the identification and recall phases but not during tracking which leads us to believe that there might be a cognitive switch between memory and attention, where the targets might not be held in working memory during tracking but rather only tracked with attention mechanisms. During recall, there is no doubt that the participant has to leverage working memory to provide the answers and even more so, the confidence level for each target. Interestingly, during the identification phase for the lateralized trials, the activity is nicely symmetrical early post-stimulus and then a change occurs and the following 500ms are not as symmetrical. We hypothesize that the period between 100-400ms engages more bottom-up brain mechanisms as a response to the stimuli and the following 500-600ms engages more top-down mechanisms to which there seems to be a difference between left and right hemispheres.

\textbf{Eye movements.}
In order to rule out the possibility that remaining eye movement artifacts be the main driver of the frontal EEG activity observed in the results, we calculated the correlation of the EOG channel pair with the lateralized signal obtained from the frontal cluster. The Pearson product-moment correlation coefficients was calculated for each participant and then averaged to obtain the group level correlation coefficient. A weak to moderate correlation is expected at the minimum, as nearby electrophysiological channels share information. For the EOG channels, we obtained a correlation coefficient of 0.765 which, when compared with the other channel pairs (F5-F6 r=0.875; F7-F8 r=0.845; AF7-AF8 r=0.823; F3-F4 r=0.798; FT7-FT8 r=0.788; AF3-AF4 r=0.769; F1-F2 r=0.739; Fp1-Fp2 r=0.581), is among the weakest correlation with the lateralized CDA-like activity presented in the results. This shows that the signal obtained isn't driven by eye movements but rather cognitive processes.

\textbf{Frequency bands.} 
During tracking, there is a strong inhibition of both delta and theta frequencies, followed by a significant reactivation of these same frequencies during recall. For such a strong switch between inhibition and activation, we ruled out the hypothesis that it could be muscle activity or artifacts based on releasing the tension after being focused during tracking, because if that was the case, we would see the opposite effect on Figure \ref{R3-ERSP-SetSizes} and the activity would spread longer with only 1 target as they would start releasing the tension and moving their body faster. However here we see a stronger but shorter activity in delta and theta during recall for 1 targets, less strong and slightly longer for 2 targets, and the weakest power but longer spread over time. This is aligned with the CDA plots in the time domain (see Figure \ref{CDA-ID-Recall}). We also ruled out the motor and speech brain-related activity from providing the answers, as such activity would be more localized and not clearly visible on all clusters as seen here.

The activation of lower frequencies (delta and theta) during recall is distributed spatially but the strong inhibition during tracking is frontal. We hypothesize that such inhibition might be coming from a top-down cognitive mechanism to ignore task distractors and prevent them from being encoded during tracking. Also, the 3D-MOT task requires the participant's full attention, because many occlusions and contacts between the objects (targets and distractors alike) are happening and one tiny lapse in attention can make the participant fail the trial. Therefore, it is of upmost importance for our top-down mechanisms to protect our attention from task-related, environment-related as well as internal distractions. Theta has been linked to memory, cognitive control and attention systems (\cite{bacsar2001gamma, karakacs2020review, eisma2021frontal}) and as observed here, has a key role in the MOT task during tracking and recall. Theta and gamma phase coupling has also been accumulating supportive evidence for playing a key role in visual processes involving working memory (\cite{sauseng2009brain, koster2014theta, lundqvist2016gamma}). This theta-gamma coupling might explain why during recall we observe also a strong gamma activation at the same time as the theta activation, strongest in the occipital region. We haven't done any phase coupling nor connectivity analysis to confirm the link between these frequencies but it is in our future plans to look at phase coupling, source localization, and connectivity. 

As for delta, the literature for its role in cognitive functions, aside from sleep studies, isn't as extensive as for the other frequency bands but it has been linked to similar attention mechanisms than theta (\cite{guntekin2016review}). Alpha, one of the most prominent rhythm in the human brain, also plays a key role in attention, especially for inhibition of distractors. A previous study providing online feedback on a 3D-MOT task based on real time alpha peak frequency has helped improve performance on the task (\cite{parsons2021enhancing}). While it is clear that alpha is playing a role, it remains unclear exactly how, as studies have shown both an increase (\cite{wutz2020oscillatory}) as well as a decrease (\cite{merkel2022electrophysiological}) in alpha during object tracking. Based on the previous evidences, looking at the grand average over the whole trial and then averaging all the trials together might be the wrong way of looking at alpha as it might be involved in a more granular fashion during a trial. Calculating the grand average might hide the subtle within-trial changes of alpha. On the ERPS graphs, we see that power changes are clearly happening around, and at, alpha frequencies however the effect isn't as strong as some other frequencies.

\begin{figure}[!htb]
    \centering
    \includegraphics[width=0.8\textwidth]{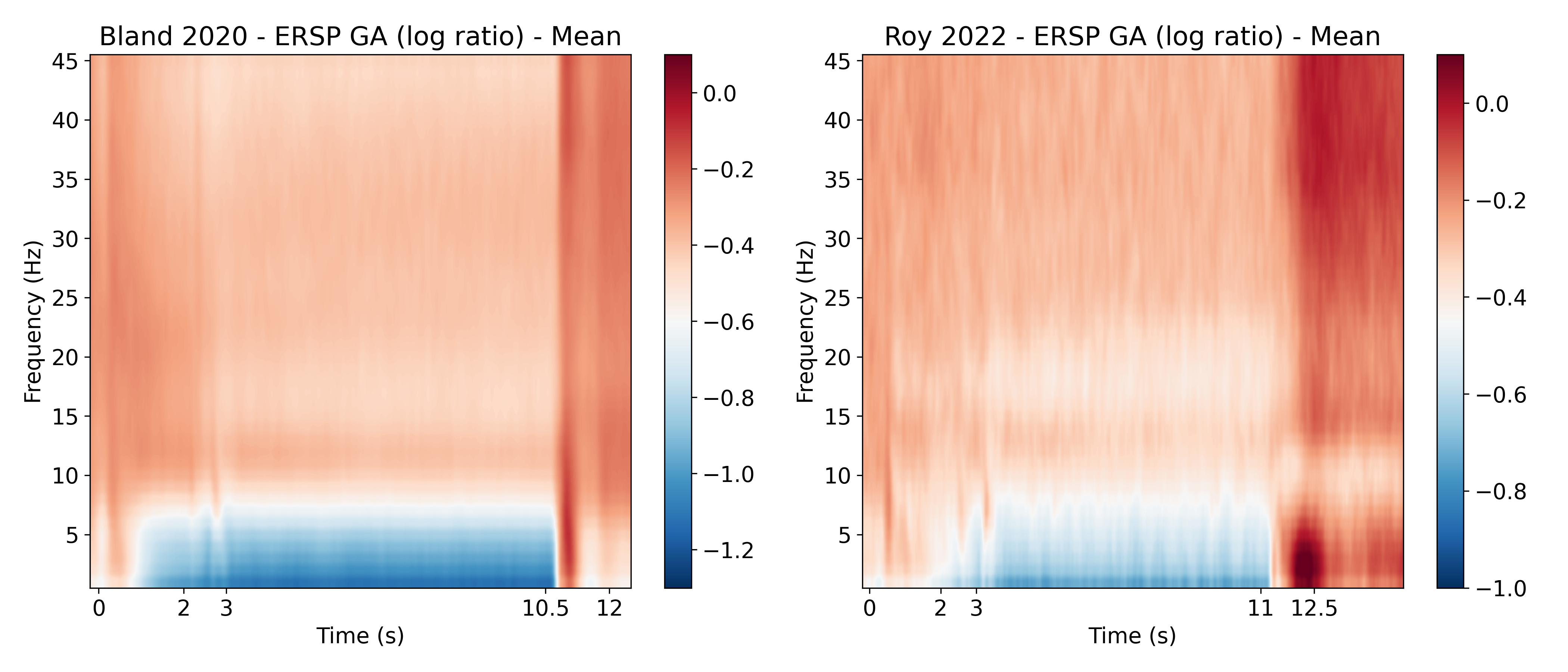}
    \caption{ERSP: Bland, 2022 vs Roy, 2022.}
    \label{D1-ERSP-Bland_vs_Roy}
\end{figure}

Given the clear EEG pattern we observed in the frequency domain for the whole trial ($\sim$15s), we wanted to compare with another MOT task and run a similar frequency analysis to see if we'd see the same pattern of activity. Thanks to Nicholas S. Bland and colleagues, who shared their EEG data from their 2020 study (\cite{bland2020gamma}), we were able to generate an ERSP graph and we found a striking resemblance. Bland and colleagues used a 2D-MOT (our task is a 3D-MOT), their stimuli were 2D circles with no filling (i.e. rings), their trials had either 2 or 4 targets presented either between-hemifield moving freely left and right but not crossing the middle part vertically or within-hemifield moving freely up and down but not crossing the middle part horizontally. At the start of the trial they presented all the circles in white, then highlighted the targets in blue for 2 seconds (like our task) after which they reverted back to white for 500ms (we used 1s in our task) before all objects start moving for 8 seconds (like our task). Their recall phase was slightly different as their participants had to click on the targets with a mouse and then received the visual feedback with the correct answers for 1.5s. In our task the participant gave their answer verbally before entering their confidence level with the keyboard and then received visual feedback with the correct answers for 2s. They recorded EEG activity with a 64 channels BrainCap (BrainProducts) device. On Figure \ref{D1-ERSP-Bland_vs_Roy}, we see the ERSP grand average across all conditions, all channels and all participants of our study on the right (as presented on Figure \ref{ERSP-GrandAverage} before) and Bland et al., 2020 on the left. We used all the trials (N=192) for all the participants (N=41). The spatial distribution on the midline was also similar to ours, showing a stronger inhibition of delta and theta in the frontal regions during tracking. Their inhibition of lower frequencies (compared to baseline) was stronger than our results so we slightly adjusted the color scale to keep a smooth range.

The main limitation of our study is the number of trials per condition. According to a recent study by William X. Q. Ngiam and colleagues on the statistical power to detect set-size effects in contralateral delay activity, it requires between 30 to 50 clean trials with a sample of 25 subjects to achieve approximately 80\% statistical power on detecting the presence of the CDA. In our study, we kept 20 subjects with an average of 45 clean (lateralized) trials after removing bad trials during preprocessing. As mentioned earlier, this experiment was part of a larger study and we unfortunately had a time limit with the participants for this task as they were doing another cognitively demanding task right after. Hence the limited amount of trials.
In the larger study, we train a machine learning classifier on 1s windows of EEG activity and therefore, the number of trials doesn't matter as much as the overall recording time.

In order to keep the participants engaged and energized (knowing they had another task after), we asked them to give the answers orally as opposed to entering them with a keyboard. Therefore, the instructor, despite entering the answers really fast on the numpad, induced an external timing factor during the recall phase. Finally, given the interesting results we observed during the recall phase, having more granular information during that phase such as the time of entry of each answer for each participant would have allowed us to investigate deeper the neural activity during that phase.

In conclusion, we believe that we offered a more holistic view of the neural substrates during a 3D-MOT task looking at the activity across the three different phases (Identification/Indexing, Tracking, Recall/Answer) both in the time and frequency domains. We also analysed the raw EEG data of another MOT study to compare our findings and observed similar overall trends, solidifying our findings. This study sheds more light at what seems to be some sort of a handoff between focused attention and working memory processes during tracking and recall and how the delta and theta bands in the frontal regions play a key role in the 3D-MOT task as they are being toggled like an on/off switch across phases.

Both the data and the code is available online: https://github.com/royyannick/3DMOT\_EEG


\section{Funding}
This work was supported by the Natural Sciences and Engineering Research Council of Canada (NSERC-RDC) (reference number: RDPJ 514052-17) and an NSERC Discovery fund.

\bibliography{main}

\section{Supplementary Material}

\begin{table}[h]
\centering
\begin{tabular}{|| l | l ||} 
 \hline\hline
\textbf{Cluster} & \textbf{Channels}  \\
\hline\hline
Frontal & Fp1, AF7, AF3, F1, F3, F5, F7, FT7 \\ & Fp2, AF8, AF4, F2, F4, F6, F8, FT8 \\ & Fpz, AFz, Fz \\
\hline
Central & C1, C3, C5, CP1, CP3, CP5, FC1, FC3, FC5 \\ &  C2, C4, C6, CP2, CP4, CP6, FC2, FC4, FC6 \\ & Cz CPz \\
\hline
Temporal & FT7, T7, TP7 \\ & FT8, T8, TP8 \\
\hline
Parietal & TP7, CP1, CP3, CP5, P1, P3, P5, P7, P9, PO3, PO7 \\ & TP8, CP2, CP4, CP6, P2, P4, P6, P8, P10, PO4, PO8 \\ & POz, Pz, CPz \\
\hline
Occipital & O1 \\ & O2 \\ & Oz, POz \\
\hline
\end{tabular}
\caption{Electrode Clusters}
\label{table_clusters-electrodes}
\end{table}

\end{document}